\documentclass[english,preprint]{revtex4-2}
\usepackage[T1]{fontenc}
\usepackage[utf8]{inputenc}
\usepackage{color}
\usepackage{prettyref}
\usepackage{amsmath}
\usepackage{amsthm}
\usepackage{amssymb}
\usepackage{graphicx}

\makeatletter

\providecommand{\tabularnewline}{\\}

\usepackage{footmisc}

{
\makeatletter
\def\frontmatter@thefootnote{%
 \altaffilletter@sw{\@fnsymbol}{\@fnsymbol}{\csname c@\@mpfn\endcsname}%
}%
\makeatother

\newrefformat{eq}{Eq.\ (\ref{#1})}
\newrefformat{fig}{Fig.\ \ref{#1}}
\newrefformat{tab}{Table.\ \ref{#1}}

\ifdefined\showcaptionsetup
 \PassOptionsToPackage{caption=false}{subfig}
\fi
\usepackage{subfig}
\makeatother

\usepackage{babel}
\begin{document}
\title{Reduced kinetic model for ion temperature gradient instability in
tokamaks with reversed magnetic shear}
\author{B. Jia}
\affiliation{Institute for Fusion Theory and Simulation, School of Physics, Zhejiang
University, Hangzhou, People's Republic of China}
\author{Q. Zhong}
\affiliation{Institute for Fusion Theory and Simulation, School of Physics, Zhejiang
University, Hangzhou, People's Republic of China}
\author{Y. Li}
\affiliation{School of Science, Tianjin University of Technology and Education,
Tianjin, People's Republic of China}
\author{Y. Xiao}
\email{yxiao@zju.edu.cn}

\affiliation{Institute for Fusion Theory and Simulation, School of Physics, Zhejiang
University, Hangzhou, People's Republic of China}
\begin{abstract}
Using the averaged magnetic drift model and a first-order finite Larmor
radius (FLR) expansion, the eigenvalue equation for the ion temperature
gradient (ITG) mode in tokamak plasmas is reduced to a Schrödinger-type
differential equation. By invoking generalized translational invariance,
the model is extended to reversed magnetic shear (RMS) configurations
and benchmarked against global gyrokinetic simulations from GTC, showing
good quantitative agreement. The analysis reveals a characteristic
double-well potential unique to RMS profiles, which gives rise to
the degeneracy between the lowest-order even and first-order odd eigenmodes
when the two potential wells are sufficiently separated radially.
The ITG instability is also found to resonate with the magnetic drift
frequency, and its maximum growth occurs when the two rational surfaces
are slightly separated. These results provide new physical insight
into ITG mode behavior under reversed magnetic shear and offer a compact,
accurate theoretical framework that bridges simplified analytic models
and global simulations.
\end{abstract}
\maketitle

\section{Introduction}

The ion temperature gradient (ITG) instability is well known as the
primary candidate for explaining anomalous ion transport in tokamaks\citep{Chen1991}.
The subsequent discovery of the internal transport barrier (ITB),
regions characterized by steep temperature and density gradients near
magnetic shear reversal points\citep{Dong1997}, has drawn considerable
attention to ITG behavior under reversed magnetic shear (RMS) configurations.
ITB formation has been observed in several major tokamaks, including
JT-60U\citep{Koide1993}, JET\citep{Kessel1994}, TFTR\citep{Levinton1995PRL},
and DIII-D\citep{Kessel1994,Lao1996,Fujita1997}. These experiments
suggest that the reversed shear region acts as an isolation layer
between the enhanced confinement region inside and the region of poorer
confinement outside\citep{Jiquan_Li_2000}. Understanding the mechanisms
that govern ITG stability in RMS configurations is therefore essential
for clarifying the physics of improved confinement.

Numerical studies of the ITG instability in RMS plasmas have revealed
several distinct features arising from the characteristic structure
of the RMS profile. Notably, radial even- and odd-parity eigenstates
with comparable growth rates have been identified in both one-dimensional
(1D) models\citep{Dong2001} and global simulations\citep{Kishimoto_2000},
in contrast to normal shear cases where even-parity ITG modes typically
dominate. These 1D slab studies attribute the appearance of such structures
to the potential well configurations intrinsic to RMS configurations\citep{jiquanli1998sheared}.
They further show that the even- and odd-parity branches may merge
when the magnetic shear gradient becomes negligibly small\citep{Dong2001}.
However, the neglect of toroidal magnetic drift in these simplified
slab models prevents direct quantitative comparison with global gyrokinetic
simulations for realistic toroidal plasmas.

In this work, we employ a newly developed ITG eigenvalue model to
elucidate the underlying stability mechanisms of ITG modes in RMS
configurations. The main challenge in formulating a 1D ITG model using
the conventional ballooning representation arises from the breakdown
of translational invariance. When toroidal coupling terms are neglected,
this difficulty can be treated in two ways: (i) by considering the
variation of the safety factor (q) profile as a second-order effect
that introduces a slowly varying envelope\citep{kim1996POP,Kishimoto_2000},
or (ii) by treating the q-profile variation as a first-order effect
using the generalized ballooning mode representation\citep{connor1978PRL,Zonca2002}.
The latter approach distinguishes the influence of the q-profile from
other global effects, such as variations in density or temperature
profiles, and is particularly advantageous for modes sensitive to
parallel structure, such as ITG.

When the toroidal coupling induced by magnetic drift is included,
however, the poloidal dependence of the drift term complicates the
formulation of a 1D kinetic toroidal ITG model. This difficulty can
be mitigated using the average magnetic drift approximation, which
retains the magnetic drift contribution while neglecting its poloidal
variation. Recent studies have shown that this approximation leads
to a Schrödinger-type ITG model\citep{BJia2025} that agrees well
with global gyrokinetic simulations from GTC\citep{ZLin1998}, underscoring
the key role of the averaged magnetic drift frequency in the ITG dispersion
relation.

By combining the generalized ballooning representation---which effectively
restores a generalized translational invariance---with the average
magnetic drift approximation, we extend the slab ITG model\citep{Dong2001}
to toroidal geometry for both normal and reversed shear cases, while
incorporating essential magnetic drift effects. This framework enables
quantitative analysis of ITG stability in RMS plasmas with improved
fidelity to global simulation results. The mode analysis under this
framework reveals a characteristic double-well potential structure
unique to RMS profiles, leading to the typical degeneracy between
even and odd eigenmodes when the potential wells are sufficiently
separated. The ITG mode is also found to be most unstable when the
two rational surfaces are slightly separated.

The remainder of this paper is organized as follows. \prettyref{sec:reducedModel}
introduces the average magnetic drift model and derives the corresponding
Schrödinger-type eigenvalue equation, along with its Weber-form representation.
\prettyref{sec:reverse-shear} extends the reduced eigenvalue model
to RMS configurations, presents numerical solutions over a broad parameter
range, and validates the results through comparison with global gyrokinetic
simulations from GTC. The potential structure of ITG modes in RMS
plasmas is examined, and the degeneracy between even- and odd-parity
modes is demonstrated. \prettyref{sec:Summary} summarizes the main
results and offers concluding remarks.

\section{Reduced kinetic model\protect\label{sec:reducedModel}}

\begin{figure}
\subfloat[]{\includegraphics[width=0.45\textwidth]{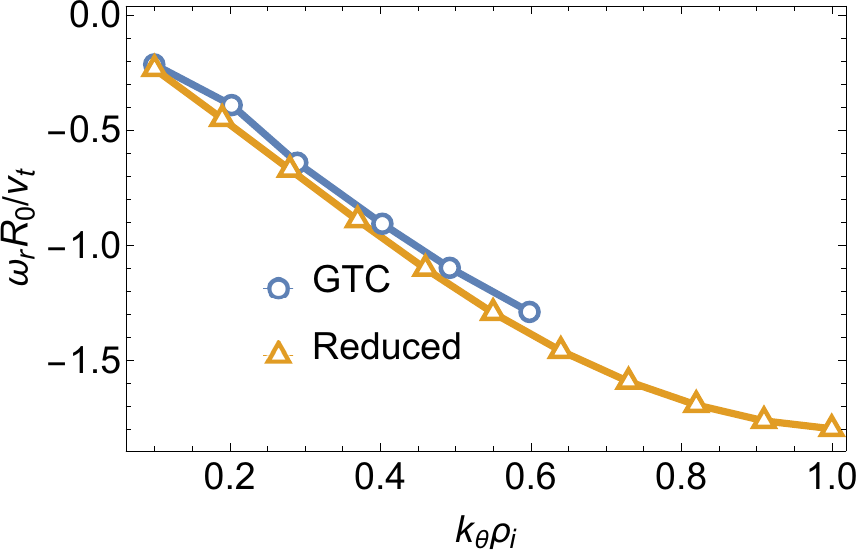}}\subfloat[]{\includegraphics[width=0.45\textwidth]{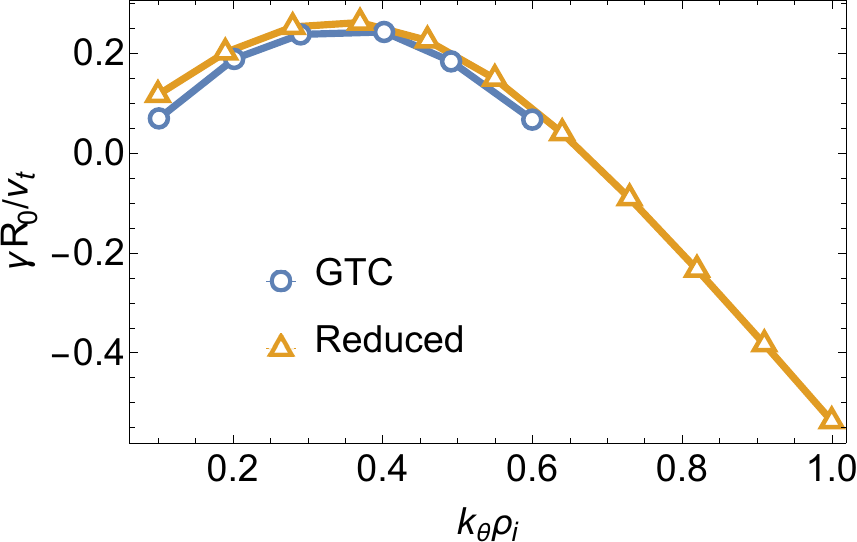}}

\caption{\protect\label{fig:FLRn}(a) Real frequency and (b) growth rate of
the ITG mode as functions of $k_{\theta}\rho_{i}$. Blue and orange
lines correspond to the GTC simulation and the reduced model results,
respectively.}
\end{figure}

We start from the ITG eigenvalue equation in the ballooning space\citep{connor1978PRL}.
Perturbed particle density $\delta n_{j}$ in the gyrokinetic theory\citep{Frieman1982}
can be decomposed into adiabatic and nonadiabatic components in the
form of
\begin{equation}
\delta n_{j}=-n_{0j}\frac{q_{j}\delta\phi}{T_{j}}+n_{0j}\int d\boldsymbol{v}h_{j}J_{0}\left[k_{\perp}\left(\eta\right)\alpha_{j}\right],\label{eq:den}
\end{equation}
where the subscript $j$ represents particle species ($j=i$ for ion,
$j=e$ for electron), $q_{e}=-e,q_{i}=Z_{i}e$ ($Z_{i}=1$), $\delta\phi$
is perturbed electrostatic potential, $n_{0j}$ is unperturbed density,
$T_{j}$ is temperature, $\alpha_{j}=v_{\perp}/\Omega_{j}$ is gyroradius
with gyrofrequency $\Omega_{j}=q_{j}B/m_{j}$, $k_{\perp}=k_{\theta}\sqrt{1+\hat{s}^{2}\eta^{2}}$
is the wave vector perpendicular to the field line with $\eta$, $\hat{s}$,
$k_{\theta}$ representing the extended poloidal angle, magnetic shear
and poloidal wave number, respectively, and the zeroth order Bessel
function $J_{0}$ corresponds to the finite Larmor radius (FLR) effects.
Electrons are assumed to be adiabatic for simplicity, i.e. $h_{e}=0$,
while, the non-adiabatic perturbed ion gyrocenter distribution function
$h_{i}$ is given by solving the gyrokinetic equation\citep{Frieman1982,Kim1993}
\begin{equation}
\left(i\frac{v_{\parallel}}{qR}\frac{\partial}{\partial\eta}+\omega-\omega_{di}\right)h_{i}=\frac{q_{i}F_{Mi}}{T_{i}}\left(\omega-\omega_{*i}^{T}\right)J_{0}\left[k_{\perp}\left(\eta\right)\alpha_{i}\right]\delta\phi\left(\eta\right),\label{eq:gk}
\end{equation}
in which various physical quantities are defined as
\begin{align*}
\omega_{*i}^{T} & =\omega_{*i}\left[1+\eta_{i}\left(v^{2}/2v_{ti}^{2}-3/2\right)\right],\\
\omega_{*i} & =T_{i}/m_{i}\Omega_{i}\boldsymbol{k}\times\boldsymbol{b}\cdot\nabla\text{ln}n_{0i},\\
\bar{\omega}_{di} & =2\epsilon_{n}\omega_{*i}=2\omega_{*i}/\left(R_{0}\text{ln}n_{0i}/dr\right),\\
\omega_{di} & =\bar{\omega}_{di}\left[\cos\left(\eta\right)+\hat{s}\eta\sin\left(\eta\right)\right]\left(v_{\parallel}^{2}+v_{\perp}^{2}/2\right)/2v_{ti}^{2},\\
F_{Mi} & =\left(2\pi v_{ti}^{2}\right)^{-\frac{3}{2}}\exp\left(-v^{2}/2v_{ti}^{2}\right),\\
\eta_{i} & =d\ln T_{i}/d\ln n_{0i}
\end{align*}
with $q$, $m_{i}$, and $v_{ti}=\sqrt{T_{i}/m_{i}}$ representing
the safety factor, ion mass and ion thermal velocity, respectively.
By substituting the density perturbations of ion and electron into
the quasineutrality condition
\begin{align}
\delta n_{i}q_{i}+\delta n_{e}q_{e} & =0,\label{eq:qn}
\end{align}
the linear ITG eigenvalue problem is formulated as:
\begin{equation}
\left(1+\frac{1}{\tau}\right)\delta\phi\left(\eta\right)=\int_{-\infty}^{\infty}d\eta^{\prime}K\left(\omega,\eta,\eta^{\prime}\right)\delta\phi\left(\eta^{\prime}\right),\label{eq:eigensys}
\end{equation}
where $\tau=T_{e}/T_{i}$ and $K$ is the velocity space integration
of the non-adiabatic response\citep{BJia2025,Romanelli1989,Dong1992,Kim1993,HsXie2017}.
\prettyref{eq:eigensys} is a nonlinear eigenvalue problem in the
form of 
\[
\sum_{n}A_{m,n}\left(\omega\right)\delta\phi_{n}=0,
\]
after discretizing the integration in the extended poloidal angle,
where $A$ is a matrix while $\delta\phi$ is a vector. It's found
that \prettyref{eq:eigensys} can be simplified by the average magnetic
drift approximation\citep{BJia2025} 
\begin{align}
\omega_{d} & \approx\bar{\omega}_{di}f\left(\hat{s}\right)\frac{v_{\parallel}^{2}+v_{\perp}^{2}/2}{2v_{ti}^{2}},\label{eq:shata-1}\\
f\left(\hat{s}\right) & =\left\langle \cos\left(\eta\right)+\hat{s}\eta\sin\left(\eta\right)\right\rangle _{-\eta_{s}}^{\eta_{s}},\label{eq:fshateq-1}
\end{align}
where operator $\left\langle \right\rangle _{-\eta_{s}}^{\eta_{s}}$
means average over the bad curvature region $\eta\in\left[-\eta_{s},\eta_{s}\right]$
with $\eta_{s}$ determined by equation $\cos\left(\eta_{s}\right)+\hat{s}\eta_{s}\sin\left(\eta_{s}\right)=0$.
Under the average magnetic drift approximation, the gyrokinetic equation
\prettyref{eq:gk} reduces to a first-order, linear ordinary differential
equation with constant coefficients. Define the radial variable $z=qRk_{\parallel}$
as the Fourier conjugate of $\eta$, and then \prettyref{eq:gk} can
be transformed into the Fourier z-space,
\begin{align}
 & \left(-\frac{v_{\parallel}}{qR}z+\omega\right)\hat{h}_{j}\left(z\right)-\bar{\omega}_{dj}f\left(\hat{s}\right)\frac{v_{\parallel}^{2}+v_{\perp}^{2}/2}{2v_{tj}^{2}}\hat{h_{j}}\left(z\right)\label{eq:gkr}\\
 & =\frac{q_{j}F_{Mj}}{T_{j}}\left(\omega-\omega_{*j}^{T}\right)\mathcal{F}\left\{ J_{0}\left[k_{\perp}\left(\eta\right)\alpha_{j}\right]\delta\phi\left(\eta\right)\right\} .\nonumber 
\end{align}
The radial ITG eigenvalue equation is given by substituting \prettyref{eq:gkr}
into the quasineutrality condition:

\begin{align}
\left\{ 1+\frac{1}{\tau}-\frac{2}{\sqrt{\pi}}\int\int dx_{\parallel}dx_{\perp}x_{\perp}J_{0}^{2}\left(\sqrt{2b}x_{\perp}\right)\exp\left(-x^{2}\right)\vphantom{\frac{\left(\omega-\omega_{*in}\left(1+\eta_{i}\left(x^{2}-\frac{3}{2}\right)\right)\right)}{\omega-\frac{\sqrt{2}zv_{ti}x_{\parallel}}{qR}-\bar{\omega}_{di}f\left(\hat{s}\right)\left(x_{\perp}^{2}/2+x_{\parallel}^{2}\right)}}\right.\nonumber \\
\left.\frac{\left\{ \omega-\omega_{*in}\left[1+\eta_{i}\left(x^{2}-3/2\right)\right]\right\} }{\omega-\sqrt{2}zv_{ti}x_{\parallel}/\left(qR\right)-\bar{\omega}_{di}f\left(\hat{s}\right)\left(x_{\perp}^{2}/2+x_{\parallel}^{2}\right)}\right\} \hat{\delta\phi}\left(z\right) & =0,\label{eq:deqdr}
\end{align}
in which, $x_{\perp}=v_{\perp}/\sqrt{2}v_{ti}$, $x_{\parallel}=v_{\parallel}/\sqrt{2}v_{ti}$
and $x=v/\sqrt{2}v_{ti}$ correspond to the normalized velocity. For
most relevant cases, $b_{r}\ll b_{\theta}$, i.e. the condition $\hat{s}^{2}\partial^{2}/\partial z^{2}\ll1$
holds in the linear operator $b=b_{\theta}\left(1-\hat{s}^{2}\partial^{2}/\partial z^{2}\right)$,
which allows \prettyref{eq:deqdr} to be solved by performing a Taylor
expansion of the Bessel function term, $J_{0}^{2}\left(\sqrt{2b}x_{\perp}\right)$,
around $b=b_{\theta}$\citep{ChenP,BJia2025}, giving the expansion
form
\begin{align}
J_{0}^{2}\left(\sqrt{2b}x_{\perp}\right) & =\sum_{n=0}^{\infty}\frac{1}{n!}\left.\frac{d^{n}J_{0}^{2}\left(\sqrt{2b}x_{\perp}\right)}{db^{n}}\right|_{b=b_{\theta}}\left(-b_{\theta}\hat{s}^{2}\frac{\partial^{2}}{\partial z^{2}}\right)^{n}.\label{eq:J0na}
\end{align}
With the first-order expansion
\begin{align}
 & J_{0}^{2}\left(\sqrt{2b}x_{\perp}\right)\approx J_{0}^{2}\left(\sqrt{2b_{\theta}}x_{\perp}\right)\nonumber \\
 & -J_{0}\left(\sqrt{2b_{\theta}}x_{\perp}\right)J_{1}\left(\sqrt{2b_{\theta}}x_{\perp}\right)\sqrt{\frac{2}{b_{\theta}}}x_{\perp}\left(-b_{\theta}\hat{s}^{2}\frac{\partial^{2}}{\partial z^{2}}\right),\label{eq:J0n}
\end{align}
\prettyref{eq:deqdr} is reduced to a Schrödinger-type second-order
differential equation for ITG eigenvalue problem\citep{BJia2025,ChenP}:
\begin{align}
\left(\frac{\partial^{2}}{\partial z^{2}}+\frac{\bar{\omega}_{di}f\left(\hat{s}\right)\left(1+1/\tau\right)+\mathcal{K}_{0}}{\sqrt{2b_{\theta}}\hat{s}^{2}\mathcal{K}_{1}}\right)\delta\phi\left(z\right) & =0,\label{eq:newModel}
\end{align}
where $\mathcal{K}_{0}$ and $\mathcal{K}_{1}$ related to the velocity
integrations are defined by
\begin{align}
\mathcal{K}_{0} & =\left[\omega-\omega_{*i}\left(1-\frac{3}{2}\eta_{i}\right)\right]\mathcal{M}_{10}-\eta_{i}\omega_{*i}\left(\mathcal{M}_{30}+\mathcal{M}_{12}\right),\label{eq:K1}\\
\mathcal{K}_{1} & =\left[\omega-\omega_{*i}\left(1-\frac{3}{2}\eta_{i}\right)\right]\mathcal{N}_{20}-\eta_{i}\omega_{*i}\left(\mathcal{N}_{40}+\mathcal{N}_{22}\right),\label{eq:K2}
\end{align}
with
\begin{align}
\mathcal{M}_{\left(n,m\right)}= & I_{nm}\left(\zeta_{\alpha},\zeta_{\beta},b_{\theta}\right),\label{eq:M}\\
\mathcal{N}_{\left(n,m\right)}= & \frac{2}{\sqrt{\pi}}\int_{0}^{\infty}dx_{\perp}\int_{-\infty}^{\infty}dx_{\parallel}\exp\left(-x^{2}\right)\label{eq:N}\\
 & \frac{x_{\perp}^{n}x_{\parallel}^{m}J_{0}\left(\sqrt{2b_{\theta}}x_{\perp}\right)J_{1}\left(\sqrt{2b_{\theta}}x_{\perp}\right)}{x_{\parallel}^{2}+x_{\perp}^{2}/2+\zeta_{\alpha}-\zeta_{\beta}x_{\parallel}},\nonumber 
\end{align}
where $\zeta_{\alpha}=-\omega/\bar{\omega}_{di}f\left(\hat{s}\right),\zeta_{\beta}=-\sqrt{2}zv_{ti}/\bar{\omega}_{di}f\left(\hat{s}\right)qR$,
and 
\begin{align}
I_{nm}\left(\zeta_{\alpha},\zeta_{\beta},b\right)= & \frac{2}{\sqrt{\pi}}\int_{0}^{\infty}dx_{\perp}\int_{-\infty}^{\infty}dx_{\parallel}\label{eq:Inm}\\
 & \frac{x_{\perp}^{n}x_{\parallel}^{m}J_{0}^{2}\left(\sqrt{2b}x_{\perp}\right)\exp\left(-x^{2}\right)}{x_{\parallel}^{2}+x_{\perp}^{2}/2+\zeta_{\alpha}-\zeta_{\beta}x_{\parallel}}\nonumber 
\end{align}
denotes the two-dimensional velocity quadratures which can be solved
by the generalized plasma dispersion function\citep{Gurcan2014,Kim1994}.
Through analytical continuation, this function effectively accounts
for wave-particle resonance, including the effects of magnetic drift.
\prettyref{fig:FLRn} (a) and (b) show the comparison of the real
frequency and growth rate of ITG solved by the gyrokinetic code GTC\citep{Rewoldt2007}
and the reduced model \prettyref{eq:newModel} for cyclone base case
(CBC), where $q_{0}=1.4$, $\eta_{i}=3.13$, $\epsilon_{n}=0.45$,
$\tau=1$, $b_{\theta}=0.32^{2}$, $\hat{s}=0.78$. It's shown that
they are consistent with each other quantitatively.

\begin{figure}[tbh]
\centering
\subfloat[]{\includegraphics[width=0.45\textwidth]{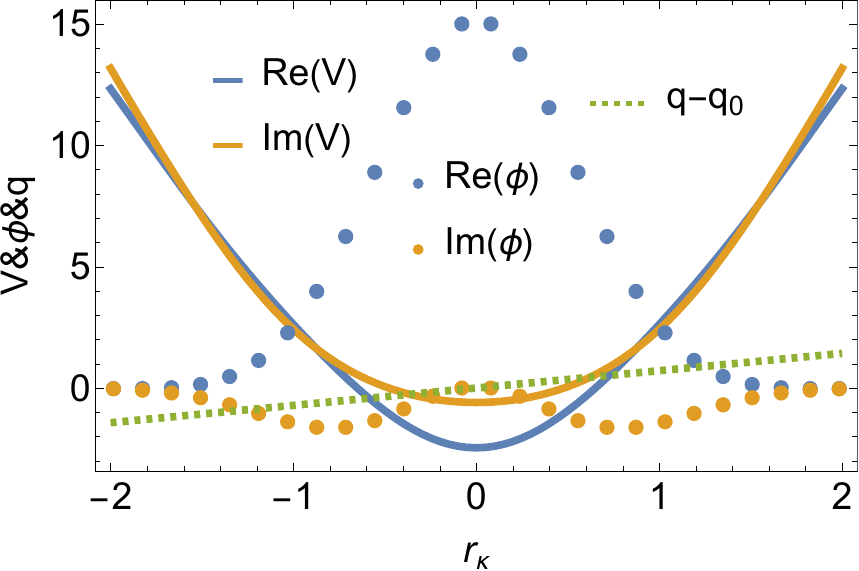}

}\subfloat[]{\includegraphics[width=0.45\textwidth]{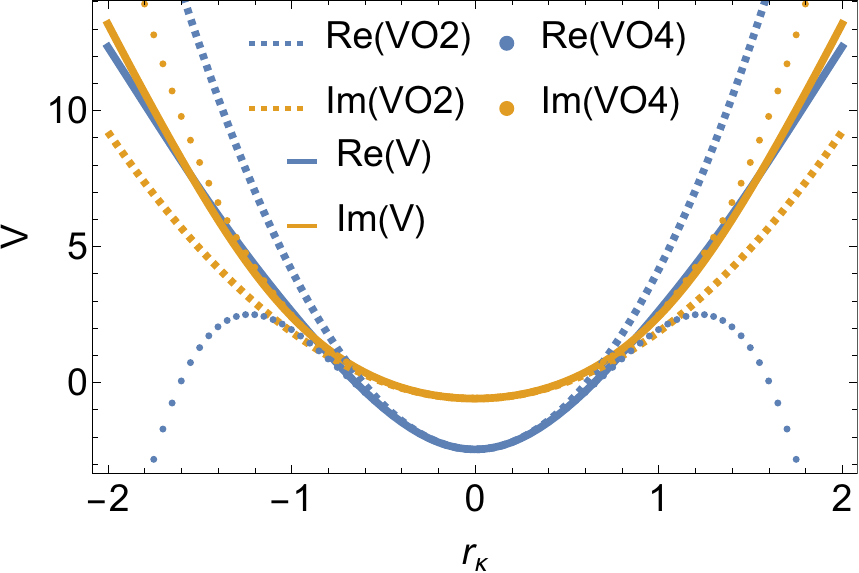}}\caption{\protect\label{fig:potentialApprox}(a) Potential well (solid lines),
mode structure (filled circles), and q-profile (green dashed line)
of the eigenvalue equation \prettyref{eq:newModel} for the CBC parameters.
(b) Full potential well (solid lines), together with the second-order
(dashed lines) and fourth-order (filled circles) Taylor expansions
of the potential well for the same parameters. Blue and orange lines
represent the real and imaginary parts of the potential well, respectively.}
\end{figure}

\begin{figure}[tbh]
\centering
\includegraphics[width=0.45\textwidth]{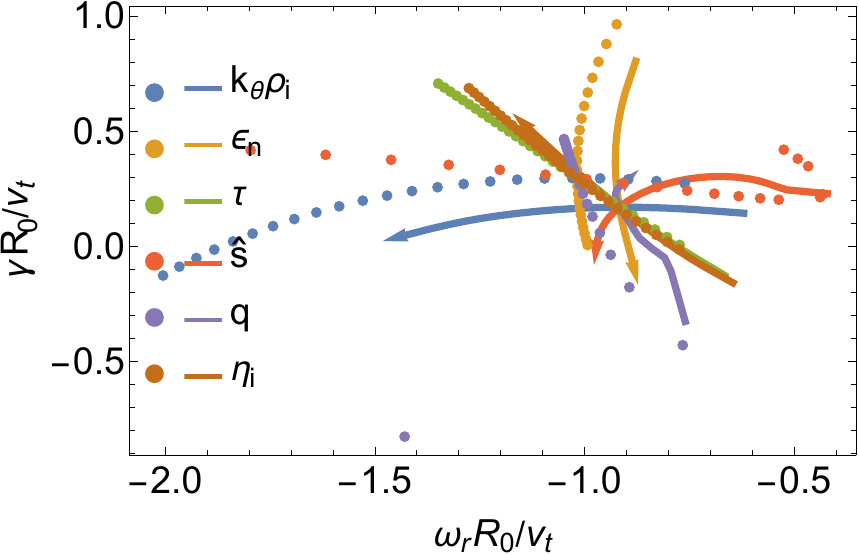}\caption{\protect\label{fig:potentialNormalShear}Eigenvalue (Real frequency
$\omega_{r}$ and growth rate $\gamma$) trajectories in the complex
plane with respect to different parameter scans. Solid lines represent
the kinetic integral model shown here as references. Filled circles
represent the Weber form equation \prettyref{eq:weberEq}}
\end{figure}

Since the reduced gyrokinetic equation \prettyref{eq:newModel} is
in Schrödinger-type, we can plot the potential well and the mode structure
in \prettyref{fig:potentialApprox} (a), given the eigenvalue solved
first. 
It's shown that the potential well is deepest at the rational surface
where the mode structure also peaks at. The potential well (V) can
be approximated by a second (VO2) or a fourth (VO4) order Taylor expansion
which is shown in \prettyref{fig:potentialApprox} (b), the fourth
order expansion exhibits qualitatively difference at large $r_{\kappa}$
which introduces non physical scattering states. Employing the second
order approximation, the simplified potential well qualitatively recover
the original potential and reduce the Schrödinger-type equation to
a Weber equation in the form
\begin{equation}
\left(\frac{\partial^{2}}{\partial z^{2}}-V\left(0\right)-\frac{V^{\prime\prime}\left(0\right)}{2!}z^{2}\right)\delta\phi\left(z\right)=0,\label{eq:weberEq}
\end{equation}
where
\[
V\left(z\right)=-\frac{\bar{\omega}_{di}f\left(\hat{s}\right)\left(1+\frac{1}{\tau}\right)+\mathcal{K}_{0}}{\sqrt{2b_{\theta}}\hat{s}^{2}\mathcal{K}_{1}}.
\]
The dispersion relation can then be formulated as
\begin{equation}
V\left(0\right)^{2}-\left(2n+1\right)\frac{V^{\prime\prime}\left(0\right)}{2!}=0;\ n\in\mathbb{Z};\ n\geq0,\label{eq:bohrCondition}
\end{equation}
where different $n$ corresponds to different energy levels. In these
models, including the original integral equation \prettyref{eq:eigensys},
the Schrödinger-type radial equation \prettyref{eq:newModel} and
the Weber equation \prettyref{eq:weberEq}, ITG eigenvalue (including
its real frequency and growth rate) is determined by six parameters,
which are $k_{\theta}\rho_{i}$, $\epsilon_{n}$, $\tau$, $\hat{s}$,
$q$ and $\eta_{i}$. The complex ITG eigenvalue varies with these
parameters, giving distinct trajectories in the complex plane as each
parameter is scanned. The solid lines in \prettyref{fig:potentialNormalShear}
represent the eigenvalue trajectories solving \prettyref{eq:eigensys}
while the filled circles correspond to that solving \prettyref{eq:bohrCondition}.
The figure demonstrates a qualitative agreement between the two approaches,
confirming the validity of the Weber form model \prettyref{eq:bohrCondition}.

\section{Reduced kinetic model in reversed magnetic shear case\protect\label{sec:reverse-shear}}

The preceding models, constructed in the ballooning space, rely on
the assumption of the translational invariance\citep{Zonca1993POFB}.
However, this assumption breaks down for the RMS configuration, where
the translational invariance is not applicable and the standard ballooning
representation fails. For the RMS configuration, the generalized ballooning
mode representation\citep{Zonca2002} or generalized translational
invariance must be employed. For simplicity, we assume a quadratic
form for the q-profile\citep{Zonca2002,Dong2001}
\begin{align*}
q & =q_{0}+\frac{\delta_{A,m}}{n}+q_{0}\hat{s}\frac{\left(r-r_{0}\right)}{r_{0}}+\frac{q_{0}^{2}s_{2}^{2}}{2}\frac{\left(r-r_{0}\right)^{2}}{r_{0}^{2}},
\end{align*}
where
\begin{align}
\hat{s} & =\frac{r_{0}q^{\prime}\left(r_{0}\right)}{q_{0}},\ s_{2}^{2}=\frac{q^{\prime\prime}r_{0}^{2}}{q_{0}^{2}},\ q_{0}=\frac{m}{n},
\end{align}
and $r_{0}$ is the reference flux surface. Therefore,
\begin{align}
nq-m & =\delta_{A,m}+\hat{s}r_{\kappa}+\frac{s_{2}^{2}}{2n}r_{\kappa}^{2}=q_{0}R_{0}k_{\parallel},\label{eq:nqMinusmReverseShear}
\end{align}
where $r_{\kappa}=k_{\theta}\left(r-r_{0}\right)$. The parameter
$\delta_{A,m}=n\left(q\left(r_{0}\right)-q_{0}\right)$ quantifies
the deviation of the safety factor $q$ at the reference flux surface
$r_{0}$ from the rational value $q_{0}$. Since $n$ and $s_{2}$
appear only in the combination $s_{2}^{2}/2n$, they may be treated
as a single parameter; however, $n=10$ is fixed hereafter for convenient
comparison with GTC results. Moreover, \prettyref{eq:nqMinusmReverseShear}
naturally reduces to the normal shear case when $s_{2}=0$, allowing
both normal and reversed shear configurations to be described within
a unified framework.

\begin{figure}[tbh]
\centering
\subfloat[]{\includegraphics[width=0.45\textwidth]{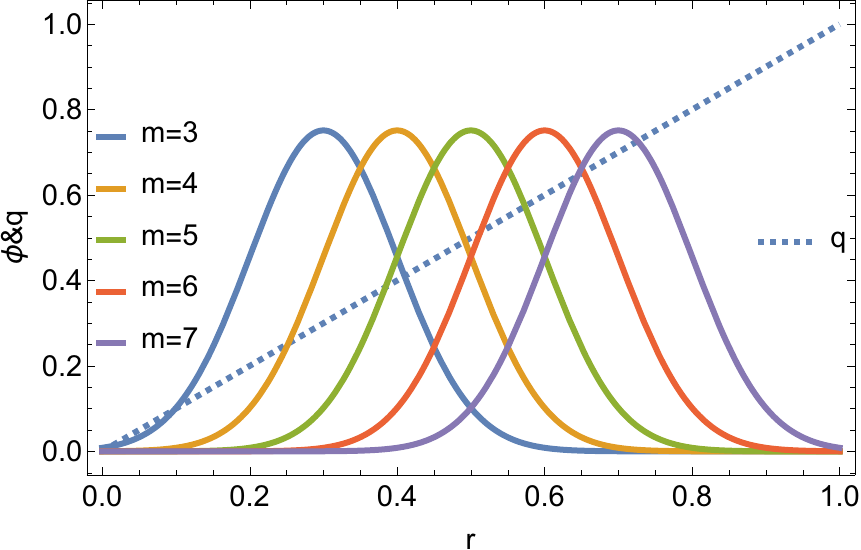}

}\subfloat[]{\includegraphics[width=0.45\textwidth]{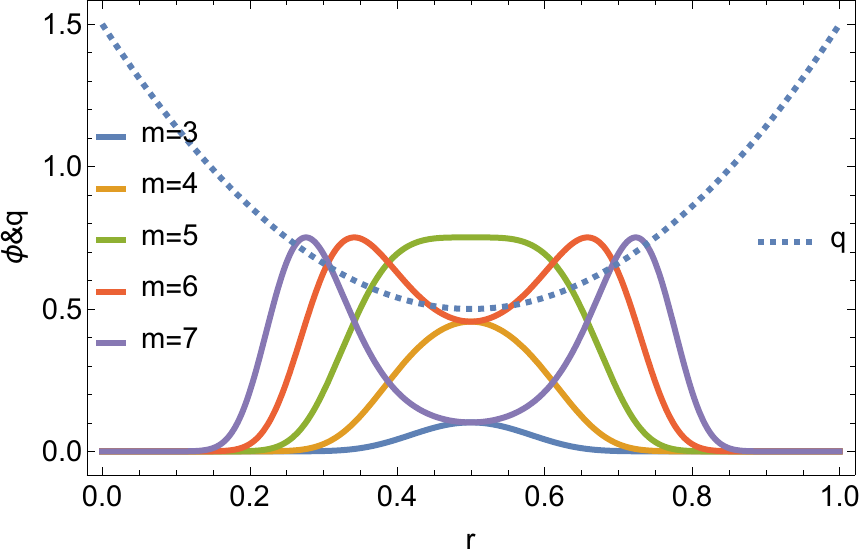}

}

\subfloat[]{\includegraphics[width=0.45\textwidth]{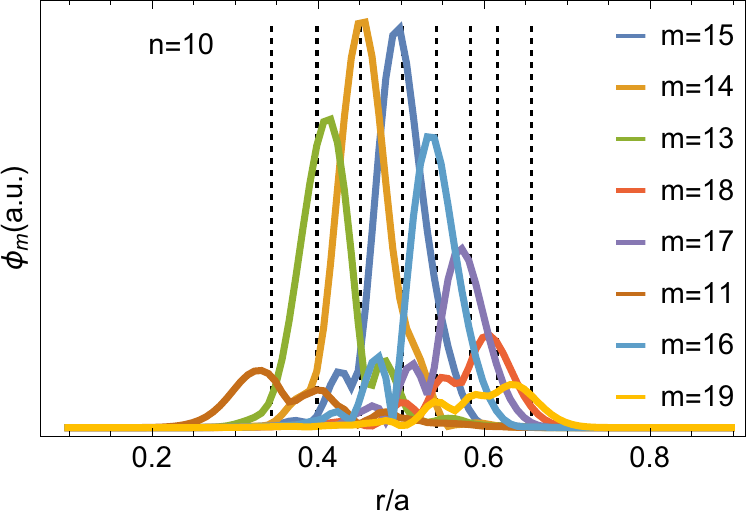}

}\subfloat{\includegraphics[width=0.45\textwidth]{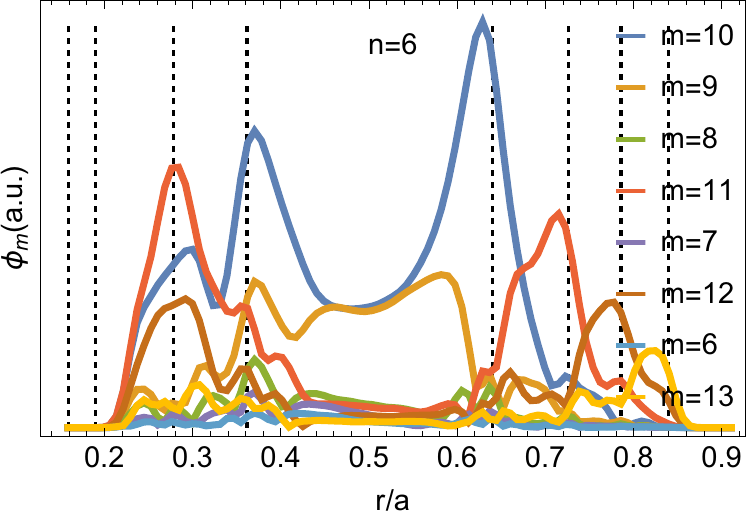}}

\caption{\protect\label{fig:modeSReverse}Radial mode structures for normal
shear (a,c) and RMS (b,d), (a,b) represent ideal translational invariance
and generalized translational invariance while (c,d) represent the
GTC simulation results correspondingly.}
\end{figure}

The ballooning representation is based on the key assumption that
the radial dependence of the mode structure arises primarily through
the parallel wave number, $k_{\parallel}=\left(nq\left(r\right)-m\right)/q_{0}R_{0}$.
In the normal shear case, where $q$ varies linearly with radius,
this leads to translational invariance. As shown in \prettyref{fig:modeSReverse}
(a), modes with different poloidal mode numbers $m$ peak at distinct
radial locations but share an identical shape, exhibiting no radial
envelope. In contrast, for the reversed magnetic shear case, the same
assumption yields the idealized mode structures shown in \prettyref{fig:modeSReverse}
(b). The behavior of modes with different $m$ then depends on the
sign of $\delta_{A,m}$:
\begin{itemize}
\item $\delta_{A,m}>0$ ($m=7,6$): modes have two peaks since there are
two rational surfaces,
\item $\delta_{A,m}=0$ ($m=5$): modes have one peak since there are only
one rational surfaces,
\item $\delta_{A,m}<0$ ($m=4,3$): modes have a smaller peak since there
are no rational surfaces.
\end{itemize}
It's shown that harmonics possessing rational surfaces share the same
maximum amplitude, so we refer to this property as generalized translational
invariance. As is shown in \prettyref{fig:modeSReverse} (c) and (d),
the GTC simulation results for the normal shear and RMS cases shows
a slight modification by a radial envelope when compared to their
ideal translational invariance counterparts. This radial envelope
may break the up-down symmetry of the 2D poloidal mode structure,
which in turn influences nonlinear transport behavior\citep{Camenen2011NF,zxlu2017POP}.
However, it is found to have a negligible effect on the linear dispersion
relation which is demonstrated in \prettyref{fig:FLRn}. 

The reduced magnetic drift model for RMS configurations is analogous
to that for normal shear cases. We adopt the average magnetic drift
model given by:
\begin{equation}
\omega_{di}=\bar{\omega}_{di}f\left(0\right)\frac{v_{\parallel}^{2}+v_{\perp}^{2}/2}{2v_{tj}^{2}}.
\end{equation}
While this work employs an average magnetic drift evaluated at the
magnetic shear reversal point, subsequent studies could incorporate
the radial dependence, treating $\omega_{di}$ as a function of the
radial coordinate $r_{\kappa}$, i.e., $\omega_{di}=\omega_{di}\left(r_{\kappa}\right)$.
The model equation for RMS configuration shows a similar structure
to the normal shear case, with the primary distinction being the specific
expression for the safety factor $q$. Consequently, the parameter
$\zeta_{\beta}$ which includes $q$, is expressed as
\begin{equation}
\zeta_{\beta}=-\frac{\sqrt{2}k_{\parallel}\left(r_{\kappa}\right)v_{ti}}{\bar{\omega}_{di}f\left(0\right)}.
\end{equation}
This formulation allows the eigenvalue equation for ITG modes to be
represented in the $r_{\kappa}$ space. The Schrödinger-type radial
eigenvalue equation can then be written as:
\begin{align}
\left(\frac{\partial^{2}}{\partial r_{\kappa}^{2}}+\frac{\bar{\omega}_{di}f\left(0\right)\left(1+1/\tau\right)+\mathcal{K}_{0}}{\sqrt{2b_{\theta}}\mathcal{K}_{1}}\right)\delta\phi\left(r_{\kappa}\right) & =0.\label{eq:finalDiffRev}
\end{align}
The finite difference discretization of \prettyref{eq:finalDiffRev}
leads to a nonlinear eigenvalue problem. This problem is solved efficiently
using the Nonlinear Inverse Iteration method\citep{Guttel_Tisseur_2017}.
Rapid and reliable convergence requires a good initial guess. Therefore,
when scanning the parameter space (e.g., the equilibrium parameter
$\tau$), the computed eigenvalue at $\tau_{i}$ is utilized as the
initial guess for the next case at $\tau_{i+1}$.

\begin{figure}[tbh]
\centering
\subfloat[]{\includegraphics[width=0.45\textwidth]{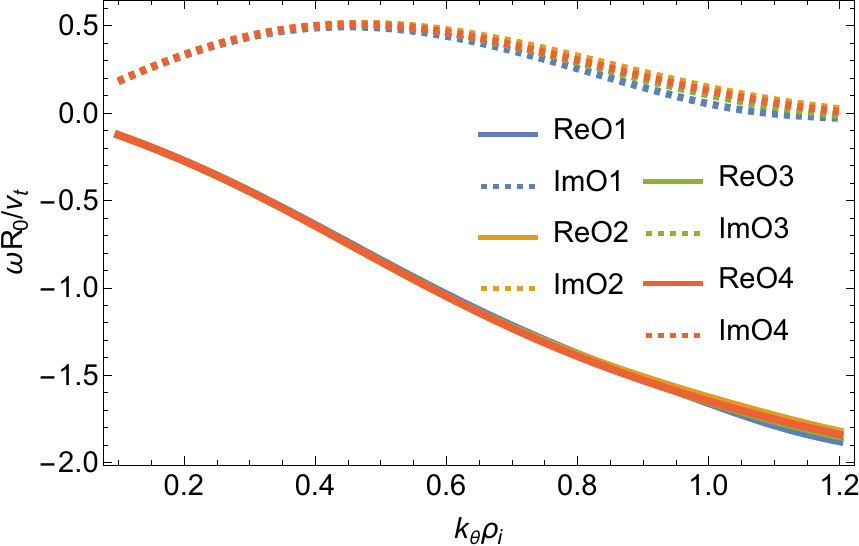}

}\subfloat[]{\includegraphics[width=0.45\textwidth]{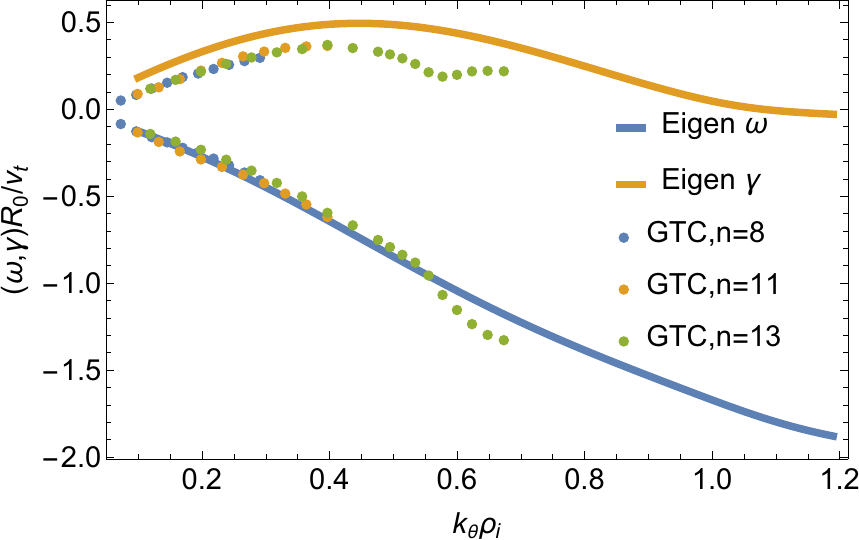}

}\caption{\protect\label{fig:GTC}(a) Real frequency (solid lines) and growth
rate (dashed lines) of the ITG dispersion relation calculated using
the first four orders of the finite Larmor radius (FLR) expansion
shown in blue, orange, green, and red lines, respectively. (b) Dispersion
relation comparison between the first-order FLR expansion and GTC
simulation results, where the GTC results for different toroidal mode
numbers (n) are represented by blue, orange, and green filled circles.}
\end{figure}

To verify the validity of the simplified model \prettyref{eq:finalDiffRev},
the CBC parameters are adjusted for the reversed shear scenario. The
equilibrium is defined by $\eta_{i}=3.13$, $\epsilon_{n}=0.45$,
$\tau=1$, $n=10$, $k_{\theta}\rho_{i}=0.4$ and a q-profile given
by
\begin{equation}
q\left(\psi_{n}\right)=2.0-3.1\psi_{n}+4.0\psi_{n}^{2},
\end{equation}
This q-profile reverses at $\psi_{n}=0.39$ (where $\psi_{n}$ denotes
the normalized poloidal magnetic flux), with a minimum of $q_{min}=1.4$.
As a function of the radial coordinate $r_{\kappa}$, the q-profile
can be approximated by a quadratic form
\begin{equation}
q=q_{0}+\frac{1}{n}\left(\delta_{A,m}+\hat{s}r_{\kappa}+\frac{s_{2}^{2}}{2n}r_{\kappa}^{2}\right),
\end{equation}
where $\delta_{A,m}=0$, $\hat{s}=0$, $s_{2}=1.78$ (these parameter
settings are hereafter referred to as the CBC-RMS parameters). The
verification is carried out in two steps. First, the convergence of
the FLR expansion is assessed by plotting the dispersion relations
for different expansion orders, as shown in \prettyref{fig:GTC} (a).
These results demonstrate excellent convergence. Next, a comparison
is made between the first-order simplified equation \prettyref{eq:finalDiffRev}
and GTC simulation results, depicted in \prettyref{fig:GTC} (b).
The results show good agreements, with minor and acceptable discrepancies
observed at large $k_{\theta}\rho_{i}$, which confirms the validity
of the simplified model \prettyref{eq:finalDiffRev} for the RMS case.

\begin{figure}[tbh]
\centering
\subfloat[]{\includegraphics[width=0.45\textwidth]{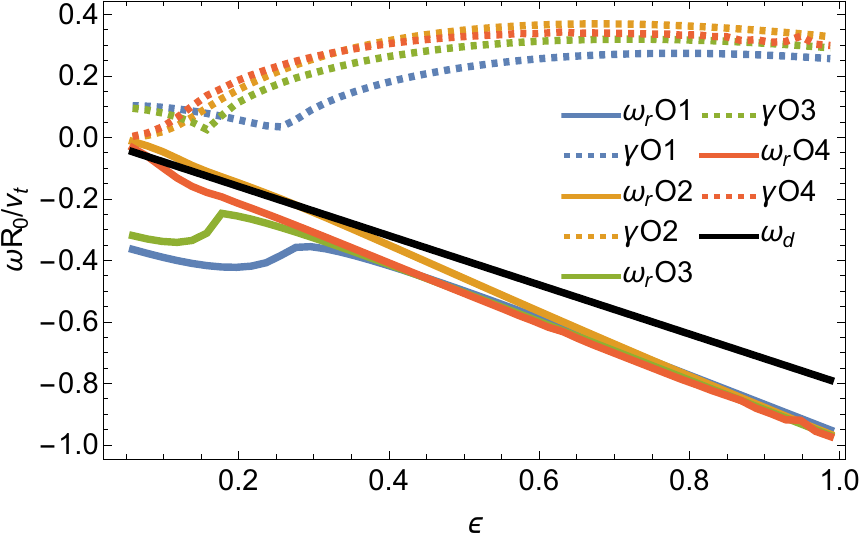}

}\subfloat[]{\includegraphics[width=0.45\textwidth]{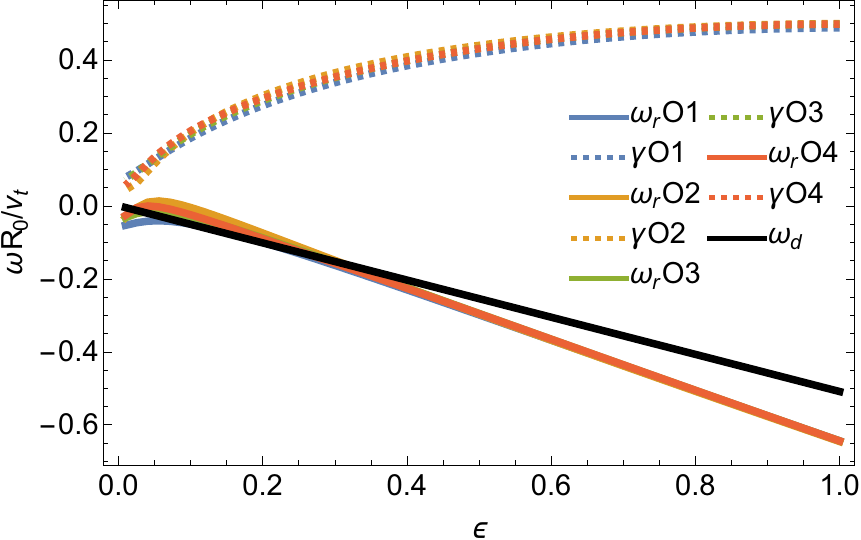}

}\caption{\protect\label{fig:zfuncdegenerate}Real frequency (solid lines) and
growth rate (dashed lines) of the ITG mode as functions of the artificial
factor $\epsilon$ for (a) normal shear and (b) reversed magnetic
shear (RMS). The blue, orange, green and red curves represent the
first four orders of the FLR expansion. The black solid line denotes
the magnetic drift frequency.}
\end{figure}

Previous numerical study suggests that the magnetic drift resonance
is important for ITG instability in normal shear configurations\citep{BJia2025},
to further analyze the resonant behavior, the average magnetic drift
frequency $\bar{\omega}_{di}$ in \prettyref{eq:finalDiffRev} is
adjusted to $\epsilon\bar{\omega}_{di}$ by introducing an artificial
factor $\epsilon$. We can then systematically investigate the influence
of magnetic drift on the frequency and growth rate using CBC or the
CBC-RMS parameters. The results of scanning parameter $\epsilon$
for normal and RMS cases are presented in \prettyref{fig:zfuncdegenerate}
(a) and (b), respectively. These figures demonstrate that the real
frequency and growth rate of different expansion orders are affected
by $\epsilon$. The resonance condition, $\omega_{r}\approx\epsilon\bar{\omega}_{di}$,
is preserved throughout the scan, confirming the important role of
magnetic drift in ITG physics. 

\begin{figure}[tbh]
\centering
\subfloat[]{\includegraphics[width=0.45\textwidth]{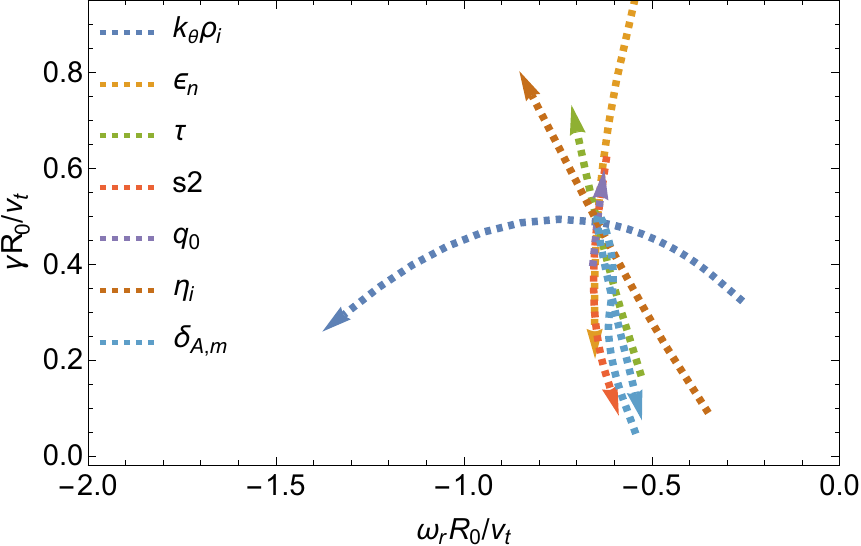}

}\subfloat[]{

\includegraphics[width=0.45\textwidth]{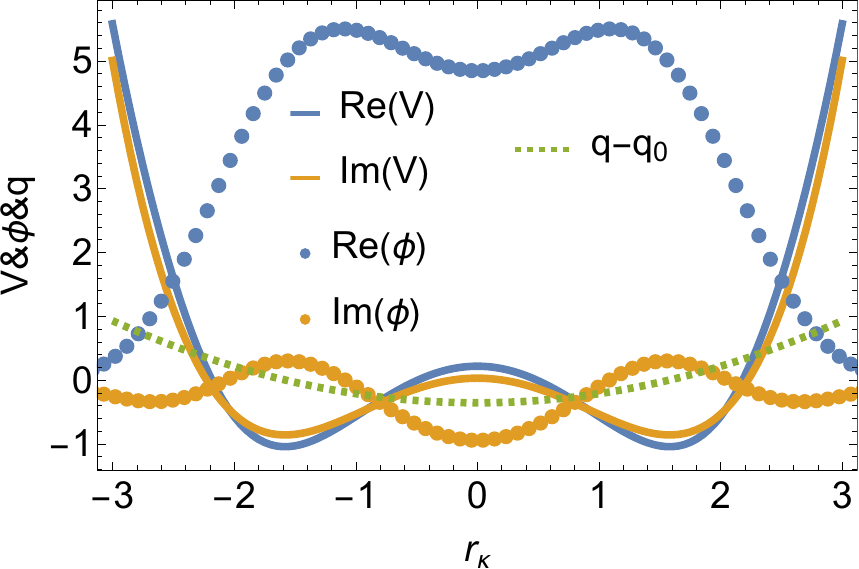}

}\caption{\protect\label{fig:reverseShearStar}(a): Eigenvalue (Real frequency
$\omega_{r}$ and growth rate $\gamma$) trajectories in the complex
plane with respect to different parameter scans in reverse shear CBC
parameters. (b): Potential well (solid lines), mode structure (filled
circles) and $q$ profile (green dashed line) of the eigenvalue equation
\prettyref{eq:newModel} in reversed magnetic shear parameters.}
\end{figure}

We also investigate the parameter dependence in RMS cases by scanning
various parameters within the ranges specified in \prettyref{tab:Parameter-scanning-ranges}.
The resulting eigenvalue trajectories, shown in \prettyref{fig:reverseShearStar}
(a), exhibit qualitative consistency with those observed in the normal
magnetic shear CBC parameters, as displayed in \prettyref{fig:potentialNormalShear}.
The results show that the growth rate increases with increasing $\tau$,
$q_{0}$, and $\eta_{i}$, and decreasing $\epsilon_{n}$. This confirms
the destabilizing character of the ion temperature gradient and the
stabilizing character of $k_{\parallel}=\left(\delta_{A,m}+\hat{s}r_{\kappa}+\frac{s_{2}^{2}}{2n}r_{\kappa}^{2}\right)/q_{0}R_{0}$
in RMS cases. The parameter $s_{2}$, one of the two new parameters
introduced in a RMS case, is found to be primarily stabilizing, similar
to the effect of $\hat{s}$ in a normal shear case. In contrast, the
other new parameter, $\delta_{A,m}$ exhibits a more complicated effect.
To explore the influence of $\delta_{A,m}$, we first plot the potential
well of the reduced kinetic model for the CBC-RMS parameters, as shown
in \prettyref{fig:reverseShearStar} (b). This potential well displays
a double-well structure, which spatially confines the peaks of the
mode structure within each well.

\begin{table*}
\centering{}%
\begin{tabular}{|c|c|c|c|c|c|c|c|}
\hline 
Parameters & $k_{\theta}\rho_{i}$ & $\epsilon_{n}$ & $\tau$ & $s_{2}$ & $q_{0}$ & $\eta_{i}$ & $\delta_{A,m}$\tabularnewline
\hline 
\hline 
Range & $\left[0.21,0.75\right]$ & $\left[0.225,0.675\right]$ & $\left[0.5,1.5\right]$ & $\left[0.89,5.34\right]$ & $\left[1.08,2.48\right]$ & $\left[1.565,4.695\right]$ & $\left[-2,0.5\right]$\tabularnewline
\hline 
\end{tabular}\caption{\protect\label{tab:Parameter-scanning-ranges}Parameter scan ranges
used in \prettyref{fig:reverseShearStar}}
\end{table*}
\begin{figure}[tbh]
\centering
\subfloat[]{\includegraphics[width=0.45\textwidth]{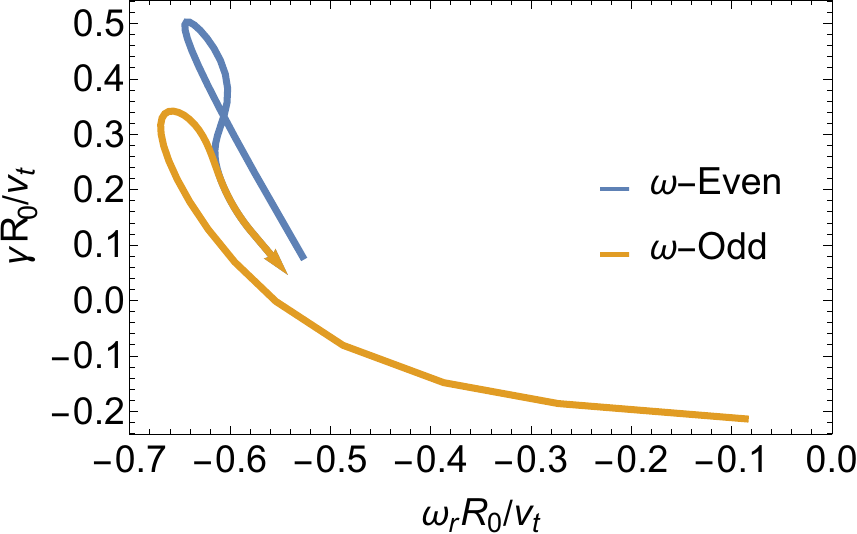}

}\subfloat[]{\includegraphics[width=0.45\textwidth]{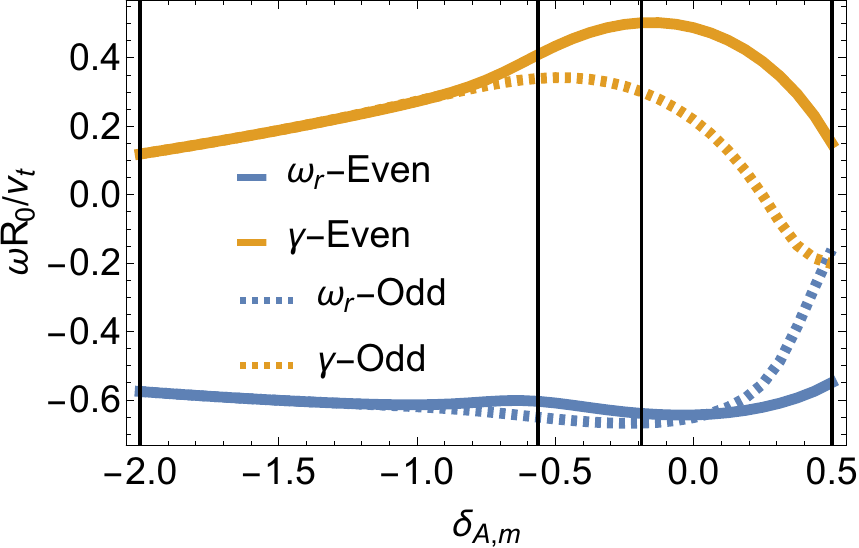}

}\caption{\protect\label{fig:EvenOddEigen}(a) Eigenvalue trajectories of the
even and odd eigenstates as functions of $\delta_{A,m}$, (b) Real
frequencies and growth rates of the even and odd modes}
\end{figure}
\begin{figure}[tbh]
\centering
\subfloat[]{\includegraphics[width=0.45\textwidth]{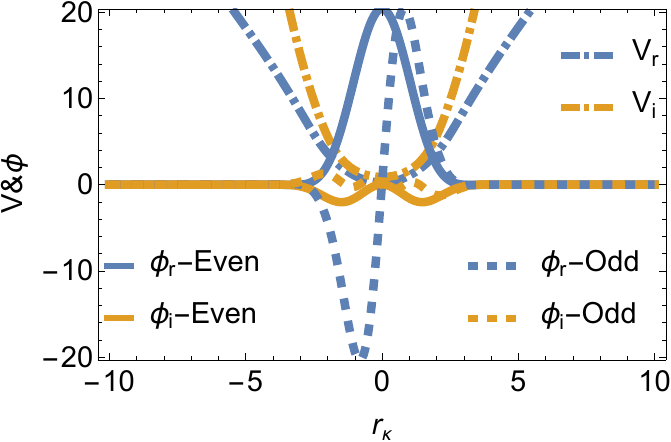}

}\subfloat[]{\includegraphics[width=0.45\textwidth]{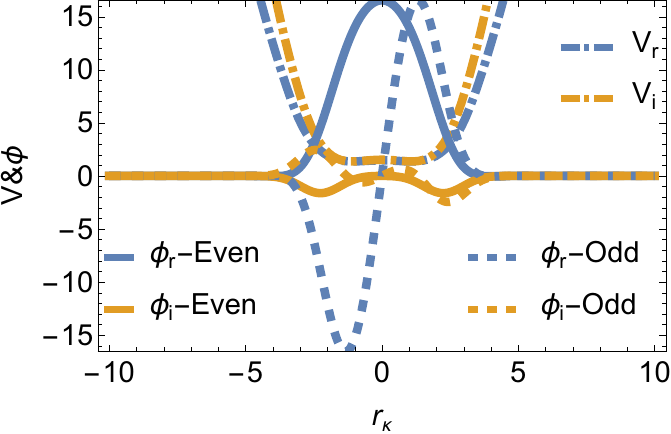}

}

\subfloat[]{\includegraphics[width=0.45\textwidth]{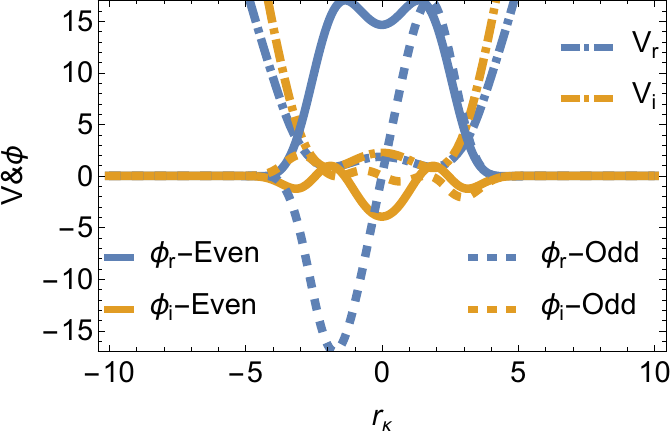}

}\subfloat[]{\includegraphics[width=0.45\textwidth]{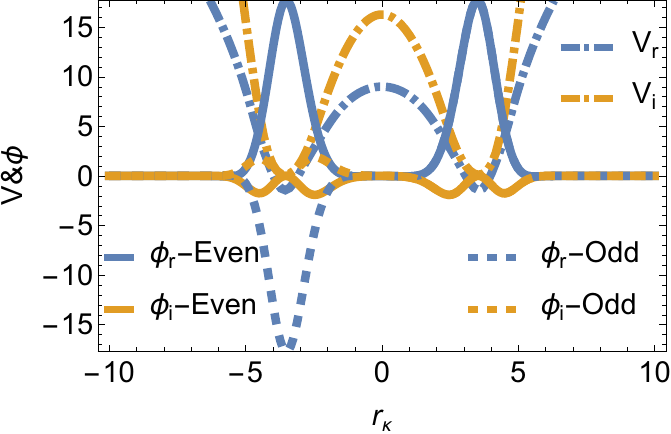}

}\caption{\protect\label{fig:eigenFunction}Eigenfunctions ($\phi$) and potential
wells ($V$) of the even and odd eigenstates for $\delta_{A,m}=$
0.5 (a), -0.185 (b), -0.5625 (c) and -2 (d). Blue and orange curves
represent the real and imaginary parts, respectively.}
\end{figure}

To illustrate the effects of the double-well structure of the potential
well, we solved the eigenvalue problem \prettyref{eq:finalDiffRev}
for different $\delta_{A,m}$. The eigenvalue trajectories of both
even and odd modes are show in \prettyref{fig:EvenOddEigen} (a),
where it is evident that their trajectories merge as $\delta_{A,m}$
decreases, with the direction of decreasing $\delta_{A,m}$ indicated
by an arrowhead. This behavior is similar to the behavior of the slab
model with RMS configuration\citep{Dong2001}. However, unlike Dong's
work where one branch disappears, we will show that the modes instead
enter degenerate states, such that both modes with even- and odd-parity
coexist when degenerate. To illustrate this, \prettyref{fig:EvenOddEigen}
(b) shows the real frequency and growth rate of both even and odd
modes as $\delta_{A,m}$ varies, The mode structures for four representative
$\delta_{A,m}$ values, indicated by the vertical black lines in \prettyref{fig:EvenOddEigen}
(b) , are presented in \prettyref{fig:eigenFunction}. These four
cases are
\begin{itemize}
\item (a) ($\delta_{A,m}=0.5$): no rational surfaces present, resulting
in a narrow eigenfunction structure.
\item (b) ($\delta_{A,m}=-0.1875$): the most unstable case, where the even
mode exhibits a single, merged peak.
\item (c) ($\delta_{A,m}=-0.5625$): the even mode peaks are partially separated,
suggesting the mode is in the process of degenerating with the odd
mode.
\item (d) ($\delta_{A,m}=-2$): the mode structure is well-separated, exhibiting
two distinct peaks, where even and odd modes degenerate with each
other completely.
\end{itemize}
When $\delta_{A,m}\geq0$, the even-parity mode exhibits a single
peak, since there are no or only one rational surface. When $\delta_{A,m}$
is smaller than 0 slightly such as $\delta_{A,m}=-0.1875$, the even
mode still has only one single peak even though two rational surfaces
are present; however, the mode structure becomes broader in this case.
As $\delta_{A,m}$ decreases further, for example $\delta_{A,m}=-0.5625$,
the single peak of even mode splits into two peaks following the separation
of the two rational surfaces. Eventually, the even and odd modes fully
degenerate when the rational surfaces are well separated, as show
in case (d) ($\delta_{A,m}=-2$). The observed degeneracy behavior
is a fundamental characteristic of the double potential well structure.
As the parameter $\delta_{A,m}$ decreases, the resulting growth of
the central potential barrier significantly suppresses the inter-well
tunneling, thereby minimizing the eigenvalue splitting and driving
the even and odd modes toward quasi-degeneracy. Consequently, the
single-peak degenerate eigenmode is also readily understood as the
linear combination (addition or subtraction) of the degenerate even
and odd modes, which is a characteristic behavior of RMS configurations
observed in the slab model too\citep{Dong2001}.

\textcolor{black}{When the even and odd modes are nearly degenerate
with each other, scanning the parameter $\delta_{A,m}$ in \prettyref{eq:finalDiffRev}
using a simple Newton iteration may fail, as the eigenvalues can switch
branches between the odd and even states. The nonlinear inverse iteration
method is employed to resolve this issue, where both the eigenvalue
and eigenvector are incorporated into the initial guess.}

\prettyref{fig:EvenOddEigen} (b) shows that the ITG mode is most
unstable when $\delta_{A,m}$ is slightly negative (corresponding
to \prettyref{fig:eigenFunction} (b)). In this regime, the mode structure
is radially broader than the case with a single rational surface ($\delta_{A,m}=0$),
reflecting the wider radial extent of region $nq-m\approx0$ and potential
$V\approx V_{min}$. When $\delta_{A,m}>0$, rational surfaces disappear,
as shown in \prettyref{fig:eigenFunction} (a), and the ITG growth
rate decreases as $\delta_{A,m}$ increases. Similarly, the growth
rate also decreases when $\delta_{A,m}$ falls below the critical
value, $\delta_{c}$, corresponding to the most unstable case (\prettyref{fig:eigenFunction}
(b)). This is because a smaller $\delta_{A,m}$ increases the separation
of the mode peaks, as shown in \prettyref{fig:eigenFunction} (c)
and (d), and enhances magnetic shear at the rational surfaces, which
has a stabilizing effect. A quantitative criterion for the most unstable
$\delta_{A,m}$ can be established by comparing the full width at
half maxima (FWHM) of the single peak mode structure with the radial
distance separating the two rational surfaces, however, that analysis
is beyond the scope of the present paper. Since parameter $\hat{s}$
does not affect the shape of the q profile, and the rational surfaces
separation can be controlled by the parameter $\delta_{A,m}$, we
keep $\hat{s}=0$ for all RMS cases in this paper.

\begin{figure}[tbh]
\centering
\subfloat[]{\includegraphics[width=0.45\textwidth]{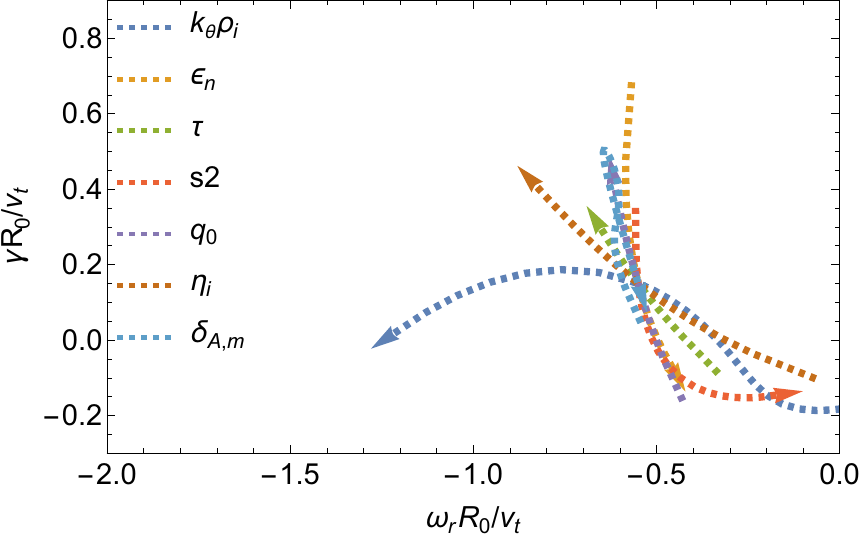}

}\subfloat[]{\includegraphics[width=0.45\textwidth]{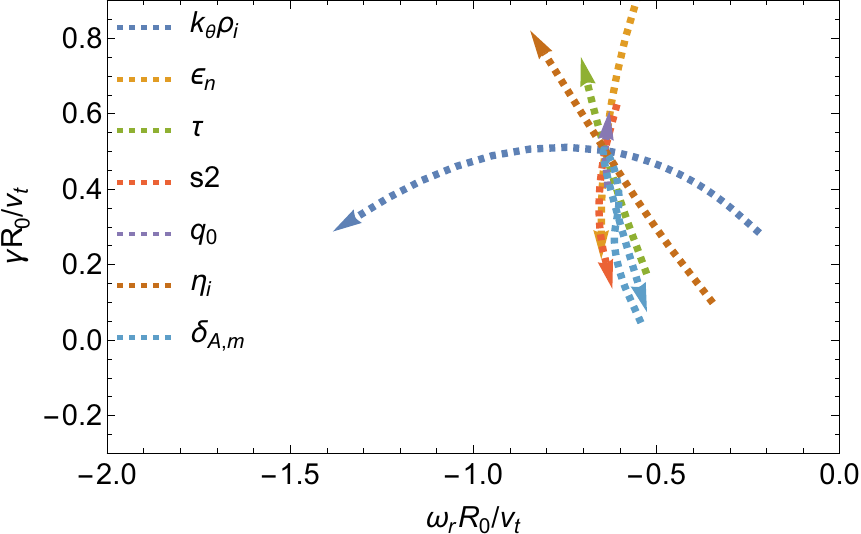}

}

\subfloat[]{\includegraphics[width=0.45\textwidth]{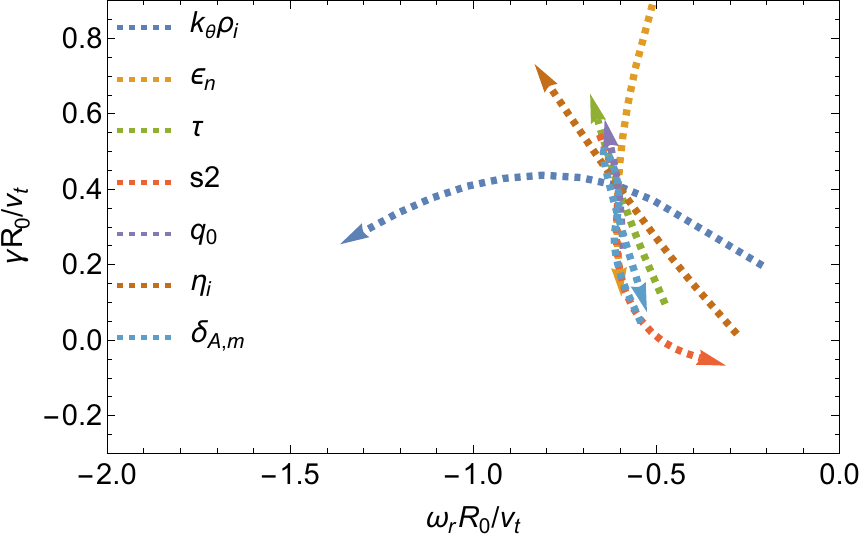}

}\subfloat[]{\includegraphics[width=0.45\textwidth]{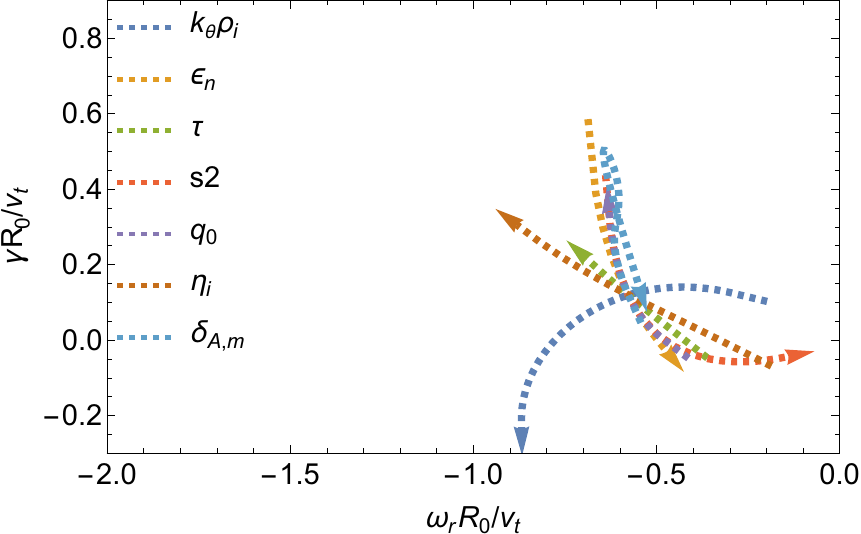}

}\caption{\protect\label{fig:eigenvalueTrajectories}Eigenvalue trajectories
as function of $k_{\theta}\rho_{i}$, $\epsilon_{n}$, $\tau$, $s_{2}$,
$q_{0}$, $\eta_{i}$, and $\delta_{A,m}$ with the base case $\delta_{A,m}$
set to 0.5 (a), -0.185 (b), -0.5625 (c) and -2 (d).}
\end{figure}

The eigenvalue trajectories shown in \prettyref{fig:eigenvalueTrajectories}
demonstrate that the fundamental parameter dependencies in the complex
eigenvalue plane are qualitatively consistent across different values
of $\delta_{A,m}$, which controls the separation between the two
rational surfaces. A notable feature in \prettyref{fig:eigenvalueTrajectories}
(d) is the counter-clockwise rotation of the trajectories, which indicates
a reduction in the destabilizing effect of ion temperature gradient,
such that may be beneficial for ITB formation according to the mixing
length estimation\citep{Dong2001}.

\section{Summary\protect\label{sec:Summary}}

In this work, we have developed a reduced kinetic model for the ion
temperature gradient (ITG) mode in toroidal geometry, applicable to
both normal and reversed magnetic shear (RMS) configurations. The
model is constructed based on the concepts of translational invariance
and its extension, generalized translational invariance, and has been
explicitly validated through comparison with global gyrokinetic simulations
from GTC. Quantitative agreement between the reduced model and GTC
results confirms the model’s reliability across experimentally relevant
parameter regimes.

Our analysis shows that the ITG mode structure and potential profile
are primarily determined by the safety factor (q) profile. In contrast
to the single-well potential typical of normal shear plasmas, the
RMS configuration gives rise to a distinctive double-well potential.
This structure leads to a characteristic degeneracy between the even
modes and the corresponding odd modes when the two potential wells
are sufficiently separated. Moreover, the ITG instability is found
to resonate with the magnetic drift frequency in both normal and reversed
shear cases.

The ITG mode exhibits maximum instability when the two rational surfaces
are slightly separated, corresponding to a small negative critical
value of $\delta_{A,m}$. Under this condition, the mode structure
becomes radially broader compared with the single-surface case ($\delta_{A,m}=0$).
Parameter scans for various RMS conditions ($\delta_{A,m}\neq0$)
further demonstrate that the results are qualitatively consistent
across configurations.

Overall, the present model provides a compact yet accurate framework
for capturing the essential physics of ITG modes in toroidal plasmas
with reversed magnetic shear. It bridges the gap between simplified
slab models and full gyrokinetic simulations, offering a valuable
tool for interpreting and predicting ITG behavior in advanced confinement
scenarios.
\begin{acknowledgments}
This work is supported by the National Magnetic Confinement Fusion
Energy R \& D Program of China under Grant No. 2019YFE03060000, as
well as by the National Nature Science Foundation of China (NSFC)
under Grant No. 12375225 and Grant No. 12405265. The computing resources
and the related technical support used for this work are provided
by the National Supercomputer Center in Tianjin and the Kylin-2 Supercomputer
at Zhejiang University.
\end{acknowledgments}

\bibliographystyle{unsrt}
\bibliography{reverseShearITG}

\end{document}